# A Facile Strategy for the Growth of High-Quality Tungsten Disulfide Crystals Mediated by Oxygen-Deficient Oxide Precursors

*Denys I. Miakota[1], Raymond R. Unocic[2], Fabian Bertoldo[3], Ganesh Ghimire[1], Sara Engberg[1], David Geohegan[2], Kristian S. Thygesen[3], and Stela Canulescu[1]*

[1]Department of Photonics Engineering, Technical University of Denmark, DK-4000 Roskilde, Denmark

[2]Center for Nanophase Materials Sciences, Oak Ridge National Laboratory, Oak Ridge, Tennessee 37831, United States;

[3]CAMD and Center for Nanostructured Graphene (CNG), Department of Physics, Technical University of Denmark, 2800 Kgs. Lyngby, Denmark

ABSTRACT

Chemical vapor deposition (CVD) has been established as a versatile route for the large-scale synthesis of transition metal dichalcogenides, such as tungsten disulfide ($WS_2$). Yet, the role of the precursor composition on the efficiency of the CVD process remains largely unknown and yet to be explored. Here, we employ Pulsed Laser Deposition (PLD) in a two-stage process to tune the oxygen content in tungsten oxide ($WO_{3-x}$) precursors and demonstrate that the presence of oxygen vacancies in the precursor films leads to a more facile conversion from $WO_{3-x}$ to $WS_2$. Using a joint study based on *ab initio* density functional theory (DFT) calculations and experiments, we unravel that the oxygen vacancies in $WO_{3-x}$ can serve as niches through which sulfur atoms enters the lattice and may facilitate an efficient growth of $WS_2$ crystals. By solely modulating the precursor stoichiometry, the photoluminescence emission of $WS_2$ can be greatly enhanced, while the size of $WS_2$ domains increases significantly. Atomic resolution scanning transmission electron microscopy (STEM) reveals that tungsten vacancies are the dominant intrinsic defects in mono- and bilayers $WS_2$. Moreover, our data reveal that grain boundaries in bilayer $WS_2$ emerge upon the coalescence of AA' and AB-oriented crystals, while turbostatic moiré patterns originate upon formation of distinct grain boundaries between the bottom and bottom layers. The atomic resolution images show local strain buildup in bilayer $WS_2$ due to competing effects of complex grain boundaries. Our study provides a means to tune the precursor composition in order to control the lateral growth of TMDs, while revealing insights into the different pathways for the formation of grain boundaries in bilayer $WS_2$.

Tungsten disulfide ($WS_2$) is an intriguing material in the layered transition metal dichalcogenides (TMDs) family. $WS_2$ can usually exist in two phases with distinct electronic properties, i.e., the 2H semiconducting phase with a trigonal prismatic geometry and 1T metallic phase with a tetragonal symmetry.[1] In particular, 2H $WS_2$ in its monolayer form is a direct band gap semiconductor, with an energy gap of ~2 eV, as demonstrated both theoretically and experimentally.[2,3] Among two-dimensional 2D TMDs, $WS_2$ monolayers exhibit a high photoluminescence (PL) quantum yield (> 6.3%), which can be further enhanced *via* doping,[4,5,6] excellent thermal stability, mechanical flexibility[7] and access to valley degree of freedom.[8] Several approaches have been exploited to obtain wafer-scale synthesis of TMD monolayers, including chemical vapor deposition (CVD),[9,10] metal-organic chemical vapor deposition (MOCVD), [11,12] molecular beam epitaxy (MBE),[13] atomic layer deposition (ALD),[14] pulsed laser deposition (PLD).[15,16,17,18] For instance,[19] conventional CVD synthesis of 2D-TMDs involves evaporation of stoichiometric tungsten trioxide ($WO_3$)[4,9,20] or molybdenum trioxide ($MoO_3$) powders.[21] The narrow optimum window of the CVD synthesis requires a precise control of the precursors vapor phase composition, chalcogen flow rate, etc. It has been shown that the conversion of $MoO_3$ to $MoS_2$ evolves via an intermediate step, during which $MoO_3$ is first partially reduced and then sulfurized into $MoS_2$.[22] Moreover, on the basis of previous findings, variation in the oxide precursor composition can greatly impact the growth of $WS_2$ flakes.[23,24]. Recently, it was demonstrated that the epitaxial oxide films grown by PLD can be used to produce quasi-crystalline TMDs with improved device performance.[19] However, despite a vast number of studies on CVD synthesis, the role of the oxygen vacancies of the precursors on the growth of TMD domains remains largely unexplored. This subject is of fundamental importance note only for the synthesis of TMDs, but also because oxygen substitution in basal plane TMDs can open new avenues for engineering of 2D electrocatalysts.[25]

In this work, we unravel how the presence of intrinsic oxygen vacancies in the non-stoichiometric tungsten oxides ($WO_{3-x}$, 0< x<1) grown by PLD leads to a more facile conversion from $WO_{3-x}$ to $WS_2$ films. A two-stage growth process was developed, which employs tunability of the oxygen vacancies in uniform $WO_{3-x}$ precursors to independently control the nucleation, lateral growth and ultimately the $WS_2$ domain size. Our study suggests that native oxygen vacancies in the PLD-grown precursors can serve as active sites through which sulfur atoms enters the lattice and facilitates the growth of $WS_2$ crystals with high photoluminescence (PL) emission and large domain size. Atomic resolution scanning transmission electron microscopy (STEM) images reveal how the single and bilayer $WS_2$ crystal orientation develops for low-temperature (nearly stoichiometric oxides) compared to high-temperature (reduced oxides) grown precursors. By solely modulating the precursor composition, the effect of precursor composition on the surface diffusion and thus lateral growth can be clearly discerned, providing insights into the parameters that control $WS_2$ domain growth.

## RESULTS AND DISCUSSION

### Pulsed Laser Deposition of $WO_{3-x}$ precursors

To synthesize $WS_2$ crystals, $WO_{3-x}$ (0< x<1) thin films and sulfur powder were used as precursors and reactant materials, respectively (see Materials and Methods). The $WO_{3-x}$ thin films were grown by PLD on (0001)-oriented $Al_2O_3$ substrates. To experimentally achieve a modulation of the oxygen vacancy content ($V_O$), $WO_{3-x}$ films were prepared at different substrate temperatures, i.e. 500 ºC, 600 ºC, 700 ºC and 800 ºC and background gases, i.e., oxygen ($O_2$) and argon (Ar). As it will be discussed next, these two parameters determine the extent to which the $V_O$ content can be tuned in the precursor films. In contrast to the CVD process, which relies on the use of oxide powder, PLD allows the growth an uniform and epitaxial film as the precursor source. In a second step, the as-grown oxide films were converted into $WS_2$ in a standard CVD furnace at a temperature of 900 ºC, which is a standard temperature for CVD synthesis of $WS_2$.[9]

We performed high-resolution X-ray photoelectron spectroscopy (XPS) to investigate the composition of $WO_{3-x}$ films. XPS data show that the $WO_{3-x}$ films grown in $O_2$ at a temperature of 500 ºC are nearly stoichiometric (Figure 1 and Table S1). The XPS spectrum of the W 4f consists of two spin–orbit doublets corresponding to $W^{6+}$ in fully coordinated $WO_3$ (W $4f_{7/2}$ binding energy 35.9 eV) and a small fraction of $W^{5+}$ in reduced $WO_{3-x}$ (W $4f_{7/2}$ binding energy 34.7 eV). Increasing the substrate temperature to 700 ºC and above, new W 4f doublets appear at binding energies of 33.4 eV and 35.3 eV due to emergence of a lower valence state of W, i.e., $W^{4+}$. The peaks at 41.7 eV and 40.4 eV were attributed to W $5p_{3/2}$. The corresponding S 2p spectra are shown in Figures S1 and S2 and the W 4f XPS spectra of $O_2$-grown precursors are shown in Figure S3.

To further modulate the $V_O$ content, $WO_{3-x}$ films were grown by ablation of the $WO_3$ target in Ar. The XPS spectra of the corresponding $WO_{3-x}$ precursors reveal the presence of three distinct valence states of $W^{6+}$, $W^{5+}$ and $W^{4+}$ corresponding to the W $4f_{7/2}$ peak at binding energy of 35.9 eV, 34.1 eV and 32.8 eV, respectively (Figure 1e-h). The evolution of the W chemical state in the $WO_{3-x}$ precursors with the change in the growth temperature and background gas is depicted in Figure 1 i,j**.** Films grown in $O_2$ at 500 ºC have a very high fraction of $WO_3$ ($W^{6+}$ fraction of ~ 0.91), along with a small fraction of $W^{5+}$ (~ 0.09). Here, $W^{x+}$ fraction, x= 4, 5, 6 denotes the ratio of the W 4f peak area (sum of the $W4f_{7/2}$ and $W4f_{7/2}$ peak areas) attributed to W in a given valence state ($W^{x+}$) to the total W 4f peak area, i.e., the sum of $W^{4+}$, $W^{5+}$ and $W^{6+}$ peaks areas. Increasing the growth temperature to 600 ºC, the $W^{5+}$ content increases to ~ 0.12. The most significant change occurs at 700 ºC, for which the fitting yields W fraction of $W^{6+}$ ~ 0.79, $W^{5+}$ ~ 0.10 and $W^{4+}$ ~ 0.11. Finally, at the growth temperature of 800 ºC, the $W^{4+}$ fraction increases to ~ 0.14. For the films grown in Ar, a significant fraction of $W^{4+}$ is already observed at 500 ºC ($W^{4+}$~ 0.41) and which increases to ~ 0.50 for oxides grown at 800 ºC.

The XPS analysis demonstrate that the oxygen deficiency ($V_O$) in $WO_{3-x}$ precursors can be greatly tuned from $O_2$ to Ar background gas, and to a degree significantly larger than reported previously.[26] The average composition of the films decreases from ~2.95 to ~2.45 from $O_2$ to Ar-grown films (see Table I). This significant variation in $V_O$ content arises from the different gas-phase reactions that occur during propagation of the plasma-plume in gas. The complex phenomena of laser ablation into gas, involving plasma-plume free expansion and shock wave propagation, has been the subject of many theoretical and experimental studies.[27,28,29,30,31] Considering first the case of pulsed laser ablation in the reactive $O_2$ gas, the plasma-plume expansion in $O_2$ leading to the formation and propagation of the shock wave favors the dissociation of the $O_2$ molecules. The interaction of plasma plume species (i.e., W atoms and ions) with oxygen will modify the plume chemistry through formation of molecular $WO^*$ excited species[27]. On the contrary, in the case of the inert Ar gas, the plasma chemistry is quantitatively different and determined by the W and O atomic species. On the basis of our findings, we can conclude that for PLD in $O_2$, the gas-phase reactions are an important pathway for the oxygen incorporation in $WO_{3-x}$, which is not the case in Ar gas.

The stability of oxygen vacancies in sub-stoichiometric oxides can be explained through their formation energy determined from density functional theory (DFT) calculations. The formation energy ($E^F$) of $V_O$ was calculated for various suboxides with compositions varying from $WO_{2.944}$ to $WO_{2.0}$ (see Figures S7 and S8). In the whole range of oxide compositions, $V_O$ was found to have the lowest formation energy for oxides close to stoichiometric $WO_3$ ($WO_{2.944}$) while the formation energy of $V_O$ increases roughly 4 times from $WO_{2.994}$ to $WO_2$. Moreover, in the whole range of O chemical potential, formation energy of $V_O$ in $WO_{2.944}$ was found to increase from 1.055 eV under an O-poor environment to 1.775 eV under an O-rich environment. This is well aligned with our experimental observations, for which a high $V_O$ content was observed in films grown under an $O_2$-poor environment and high temperature.

While the theoretical calculations do not account for the stability of the aforementioned oxides, they suggest that oxygen vacancies in $WO_{3-x}$ can easily form under O-poor environment. In particular, oxygen vacancies are a signature of oxide films grown by PLD[32] and can cause an atomic re-arrangement of the $WO_3$ crystal to compensate for the high oxygen deficiencies.[33] Yet, the detailed mechanism of surface reconstruction in sub-stoichiometric tungsten oxide is beyond the scope of the work. A wide range of oxides can form between $WO_3$ and $WO_2$. Among all, $W_{18}O_{49}$ ($WO_{2.72}$) and $W_{20}O_{58}$ ($WO_{2.9}$) are the most thermodynamically stable phases, and which exhibit metallic characteristics as the Fermi level moves up in the conduction band as compared to the parent $WO_3$[33,34]. It is therefore likely that the aforementioned phases either co-exist in the PLD-grown films, or metastable oxides present in the PLD precursors transform into most thermodynamically stable phases during the CVD process. Nanostructures of $WO_{2.72}$ were previously synthesized by various approaches, such as heating a W foil at high temperature (1600 °C)[35], photochemical reduction of carbon dioxide by visible light[36], thermally evaporating $WO_3$ powders, electron-beam induced synthesis[37], etc.

The X-ray diffraction (XRD) patterns shown in Figure 1k and Figure S4 reveal that monolithically distorted rutile (020)-oriented $WO_2$ films can be grown on (0001) $Al_2O_3$ substrate.[38] The (020) reflection of $WO_2$ develops from a very broad to a sharp (020) oriented peak, indicating an optimum window growth of epitaxial-oriented $WO_2$ between 600 ºC and 700 ºC. The b-axis oriented growth of $WO_2$ are similar to that of monoclinic $MoO_2$.[38,39] The $WO_2$ reflection of the film grown at 800 ºC shifts significantly from 36.8º to 37.4º which can probably be assigned to a change the a (002)-preferred oriented $WO_2$[38].

**Conversion from $WO_{3-x}$ to $WS_2$ crystals**

The W 4f XPS spectra of $WS_2$ films obtained upon sulfurization of $WO_{3-x}$ precursors grown in Ar are shown in Figure 2. The W 4f XPS spectra of the sulfurized films can be tailored using three doublets, in accordance with different valence states of W (see also Tables S2 and S3 for detailed W 4f and S 2p spectra peak fitting). A similar trend was observed for the precursors deposited in $O_2$. The W 4f XPS spectra of $WS_2$ exhibit a dominant $W^{4+}$ doublet (W $4f_{7/2}$ binding energy at ~32.6 eV, full width of half maximum (FWHM) ~ 0.7 eV) which corresponds to the hexagonal 2H-$WS_2$ phase[9] with an estimated S/W ratio of ~2. In addition, residual $WO_3$ is evidenced by the presence of the W4f doublet with $W^{6+}$ valence state (W $4f_{7/2}$ binding energy at ~ 36.0 eV).[40,41] We find that the precursors grown at 500 ºC in Ar and sulfurized at 900ºC have a fraction of $WO_3$ of ~ 0.34 (see Table 1). In contrast, only an oxide fraction of ~ 0.20 was estimated in $WS_2$ based on 800 ºC-grown precursors, indicating a nearly complete conversion from oxide to sulfide. In addition, a low-intensity $W^{x+}$ doublet (binding energy W $4f_{7/2}$ at ~ 32.0 eV) is observed in the XPS spectra. Components at lower binding energies with respect to the hexagonal $WS_2$ phase can be either assigned to the 1T-$WS_2$ metallic phase[42,43] or sulfur-deficient films.[42] We hypothesize that the doublet appears due to sulfur-deficient $WS_2$ which may arise from sulfur-deficient grain boundaries, as well as various intrinsic point defects in pristine $WS_2$[42]. We note, however, that the sulfur-deficient $WS_2$ fraction is very low (~ 0.08-0.06) and any variations are within the measurements accuracy (see Table 1). Lastly, the broad contributions at ~ 38.6 eV and ~ 41.6 eV (light green-colored curves in Figure 1) are assigned to the W $5p_{3/2}$ peaks.[9,40,41] Since neither S-O bonds nor W-S-O bonds were observed in the XPS spectra (see S 2p XPS spectra in Figure S1), we can conclude that, within the sensitivity of the XPS, the intermediate $WO_xS_y$ phase is not present in the sulfurized films for Ar-grown $WO_{3-x}$ films.

Our experimental results indicate a more facile sulfurization for oxide precursors with of a high $V_O$ content. The sulfurization of $WO_3$ in the presence of $H_2S$ has been studied extensively in the past[44] and it was show that the reduction of W valence state ($W^{6+}$ to $W^{5+}$) is the first step in the sulfurization process. In addition, higher bond energy of W–O (~670 kJ/mol) compared to Mo–O (~560.2 kJ/mol) requires a higher temperature of conversion for $WS_2$ as compared to $MoS_2$ (typically below $800^0$C when using PLD-grown $MoO_x$ films as precursors, not shown). Pathways to reduce the sulfurization temperature were demonstrated via the formation of atomic H in plasma discharge which facilitates W reduction[45].

To explore the thermodynamics of the process, we have performed DFT calculations to determine the energy required to incorporate S atoms into the bulk $WO_3$ crystal (see Methods and Figure S9 for details). Our calculations indicate that the addition of interstitial sulfur ($S_i$) in $WO_3$ requires a formation energy of 3.44 eV. On the other hand, the formation energy for sulfur atom on an oxygen vacancy site ($S_O$) is only 2.55 eV; hence, incorporation of S atoms into the lattice is more energetically favorable in the presence of oxygen vacancies as compared to the interstitial sites. Next, we have extended the modelling to investigate the energetics of the sulfurization process on the $WO_3$ (001) surface simulated by a slab model. Our theoretical calculations reveal that adsorbing a sulfur atom at the oxygen site ($S_O$) site yields a negative adsorption energy of -1.256 eV. This suggests that sulfur adsorption on the $WO_3$ (001) surface in the presence of an oxygen vacancy is highly energetically favorable. This allow us to conclude that oxygen vacancies in $WO_{3-x}$ can serve as niches through which sulfur atoms enters the lattice and trigger an efficient conversion of $WO_{3-x}$ into $WS_2$.

PL and Raman spectroscopy were used to examine the optical properties of $WS_2$. The PL spectra of $WS_2$ acquired using a 532 nm laser excitation wavelength are shown in Figure 3. Remarkably, the PL spectra differ in peak intensity, peak position and FWHM as a function of the growth temperature of the $WO_{3-x}$ precursors. The overall PL emission of $WS_2$ increases ~40 fold with increasing precursor growth temperature from 500 °C to 800 °C, whereas the FWHM of the PL peak decreases gradually from the 500 °C to 800 °C-based precursors (see Figure 3a, c, e, g**)**. This suggests that the optical quality of $WS_2$ is higher for high temperature-based precursors as compared to the low-temperature counterpart. All peaks marked with asterisk are associated with the sapphire substrate.

To better understand the PL spectra, we consider the contribution of neutral exciton (X), negative exciton or trion ($X^{-}$), bi-exciton (XX), and defected-mediated exciton (D) to the overall PL emission of $WS_2$.[46,47,48] The optimal peak fitting was achieved with 100% Lorentzian functions. The PL spectra of $WS_2$ is dominated by the neutral exciton emission (A exciton), which shits from 2.02 to 1.99 eV with increasing precursor growth temperature from 500°C to 800°C. We also find that the spectral weight of the neutral X exciton decreases. The increase in the A exciton emission along with the decrease of the FHWM of the peak indicates a lower presence of defects in the high temperature-grown samples.[4] Moreover, we observe the emergence of a low-energy emission feature (D), which is attributed to excitons bound to defects.[49] Similar sub-band gap PL features were observed upon ion bombardment of 2D-TMDs. [49,50]

The Raman spectra of $WS_2$ display a Raman peak at ~178 $cm^{-1}$ which corresponds to the longitudinal acoustic LA(M) mode, a first-order Raman active mode of the LA phonons at the edge of the Brillouin zone (the M point)[51]. The dominant Raman feature at ~353 $cm^{-1}$ can be deconvoluted into three distinct contributions corresponding to the $E_{2g}^1$ (M) in-plane mode at ~347 $cm^{-1}$, 2LA(M), second-order Raman mode of the LA phonons at ~354 $cm^{-1}$ and $E_{2g}^1$ (Γ) in-plane vibration mode of the W and S atoms at ~357.5 $cm^{-1}$, in a good agreement with previous work[52]. In addition, the $A_{1g}$(Γ) peak at ~ 419 $cm^{-1}$ corresponds to the out-of-plane vibration of the S atoms in $WS_2$. The separation between the $A_{1g}$(Γ) and $E^1_{2g}$(Γ) Raman modes (denoted as Δk in Table II) is commonly used to identify the number of layers, whereas Δk of ~ 60 - 61 $cm^{-1}$ denotes a monolayer $WS_2$[51,52]. Δk values of 60.7 $cm^{-1}$, 61.1 $cm^{-1}$, 61.3 $cm^{-1}$ and 61.6 $cm^{-1}$ were obtained for the $WS_2$ films grown using 500 ºC, 600 ºC, 700 ºC and 800 ºC Ar-deposited $WO_{3-x}$ precursors, respectively (see Table 2). The data indicate that the sulfurized films consist predominantly of mono-bilayers $WS_2$. A detailed discussion of the Raman peak assignments is given in the Supporting Information.

One remarkable difference is that the $I_{2LA(M)}$/ $I_{A1g(\Gamma)}$ ratio is the lowest for $WS_2$ obtained from Ar-deposited $WO_{3-x}$ precursors at 500 ºC (see Table 2). Small variations in the $I_{2LA(M)}/I_{A1g(\Gamma)}$ ratio can arise from multilayer regions. We observe, however, that at the excitation wavelength of 532 nm, the $A_{1g}$(Γ) peak is more enhanced for the 500 ºC-based precursors. As it will be discussed next, this trend cannot be associated with resonance-enhanced Raman scattering. For monolayer $WS_2$, the first and second order Raman peaks (2LA(M), $E^1_{2g}$(Γ) and $A_{1g}$(Γ)) can be greatly enhanced for photon energies corresponding the three optical exciton excitations, commonly labeled as $X_A$, $X_B$ and $X_C$[52]. In particular, the $A_{1g}$(Γ) peak is enhanced at ~2.4 eV which corresponds to the $X_B$ exciton energy (B exciton)[52]. The B exciton energy ($E_B$) can be calculated as $E_A + \Delta_{SO}$, where $E_A$ is the A exciton energy and $\Delta_{SO}$ is the spin-orbit splitting energy. Considering $\Delta_{SO}$ of 0.39 eV for $WS_2$[53], $E_B$ was estimated to vary between ~2.408 and ~2.388 eV for the 500 and 800 ºC-based precursors. As the photon energy of 2.33 eV (532 nm) is far from the $X_B$ exciton energy, we can conclude that the enhancement of the $A_{1g}(\Gamma)$ peak due to resonance Raman scattering is negligible. An enhancement of $A_{1g}(\Gamma)$ mode (as well as the LA(M) mode) can be observed in samples with high density of defects[52,54]. We can therefore conclude that $WS_2$ monolayers based on 500 ºC Ar-deposited $WO_{3-x}$ exhibit a higher amount of defects as compared to the high-temperature counterpart. Previous findings have shown a clear correlation between the FWHM of the $A_{1g}(\Gamma)$ mode, the doping level and the presence of defects[52,54]. For $WS_2$ synthesized in this work, the FWHM of $A_{1g}(\Gamma)$ mode increases with increasing the precursor growth temperature (see Table 2) and thus can be attributed to the reduced amount of defects.

Atomic force microscopy (AFM) images reveal that the size and shape of the $WS_2$ crystals varies significantly as a function of the nature of the precursors. $WS_2$ domains with triangular-shaped grains of ~ 200 nm in size are observed for films obtained from precursors grown at 500 ºC (Figure 4a). $WS_2$ consists primarily of monolayers along with a smaller contribution from bilayers (Figure 4c). Similar morphology was observed for $MoSe_2$ monolayers grown by PLD in vacuum[55]. For the specimens based on the 700 ºC-precursors, the $WS_2$ crystals become more circularly-shaped, of similar grain size of ~200 nm, while the average surface coverage of the monolayer increases as compared to the low-temperature counterpart (see Figures S5 and S6). Finally, increasing the precursor growth temperature to 800 ºC, large $WS_2$ domains can be identified. Figure 4 shows AFM images of $WS_2$, where large islands consisting of monolayer and multilayer patches are visible. At the edge of the sample (Figure 4 e,f), isolated triangles have edge lengths ranging between 18 and 26 μm. Away from the edge of the film, i.e., in regions where the precursor films thickness increases (nucleation density is high), islands merge into a quasi-continuous monolayer, and multilayer regions emerge, while the $WS_2$ crystal size decreases (Figure S5).

The atomic structure of $WS_2$ was studied using atomic-resolution annular dark-field (ADF) imaging using an aberration-corrected scanning transmission electron microscope (STEM). For this purpose, the $WS_2$ specimens grown on sapphire were transferred onto a TEM grid, as described in Methods. Figure 5 shows the low-resolution STEM images of the as-transferred $WS_2$ films based on 500 and 800 $^{0}$C-grown oxide precursors. The intensity of the STEM image is directly related to the atomic number of the atoms (Z) and with the number of atomic layers.[15,56] The atomic structure of 2H $WS_2$ can be unambiguously distinguished in Figure 5d, in which sulfur atoms ($Z$ = 16) have much lower intensity than tungsten atoms ($Z$ = 74). The most common point defects identified in monolayer $WS_2$ are single W vacancy ($V_W$) and double W vacancies ($V_{2W}$) (Figure S11), in good agreement with previous work.[57] These intrinsic defects are observed for all specimens, regardless on the precursor synthesis route. We also note that these intrinsic point defects are significantly different from one-step growth PLD[15] or sputtering,[58] where antisite defects are dominant intrinsic defects. This suggests that the $WS_2$ domains grow under S-rich environment.[59,60] While the AFM micrographs can only provide a microscale picture of the TMD domains (Figure 4 a,c), atomic resolution images reveal the presence of triangular-shaped bilayer $WS_2$ domains for all samples, regardless of synthesis conditions. Lastly, the XPS data reveal a high oxygen content in the sulfurized films in the form of oxide and to a lesser fraction as $WO_xS_y$. The presence of the oxygen atoms in $WS_2$ films could not be accurately assessed by means of STEM as the relatively weak intensity contrast makes it difficult to distinguish between the sulfur on oxygen site ($S_O$) and S vacancy ($V_S$).[61]

$WS_2$ bilayers obtained from 800 $^0$C precursors are found to have predominantly the 2H (AA$^{'}$)/3H (AB) stacking (Figure 5b). In the case of bilayer $WS_2$ based on 500 $^0$C-grown precursors, the atomic structure is more complex. Grain boundaries (GBs) formed by atomic stitching of randomly oriented domains lead to the formation of bilayers with AA$^{'}$/AB stacking, as well as turbostatic moiré patterns. Depending on the precise manner the bilayer meet, a wide variety of dislocation cores and grain boundaries can form. Figure 6 shows representative high-resolution images of bilayer $WS_2$ with various GBs. When the bottom and top layers share the same GB, here referred to as *bilayer* GB, the $WS_2$ bilayers have a 2H (AA$^{'}$)/3H (AB) stacking orientation (Figure 6a). In this case, a monolayer forms by atomic stitching of domains with a tilt angle of 23.5$^0$, resulting in dislocation rings of different sizes, such as single and double W clusters consisting of aligned 8-fold dislocation rings, in agreement with previous reports.[15] An atomic resolution image of a monolayer $WS_2$ with dislocation cores at the GB is shown in Figure S12. In the second stage of the growth, top layers nucleate on individual domains maintaining a similar orientation of the original bottom grains (AA$^{'}$ and AB). They crystals continue to expand laterally until they reach the GB, when the growth stops. A schematic of the GB in bilayer $WS_2$ is shown in Figure 6g, *left-hand* sketch. Interestingly, we find that the on the length scale of several nanometers, the AA$^{'}$ stacked bilayer has undergone an *in-plane* translation, likely due to the build-up local strain near the GB (Figure 6b). In addition to bilayer domains sharing the same GB, bilayers $WS_2$ with distinct GBs for the bottom and top layers are observed (Figure 6d,e), resulting in the emergence of moiré patterns. In this case, in the first stage of the growth, low (7.5$^0$) and high-angle (20.5$^0$) monolayer GBs are formed. Similarly, top layers nucleate on grains A and B with AB/AA$^{'}$ crystal orientation. The top layer which nucleates on grain A extends over the low-angle GB (7.5$^0$) on the neighboring grain resulting in the formation of a turbostatic moiré pattern (bilayer C). On the contrary, the top layer which nucleates on grain B stops at the high-angle GB. We hypothesize that the formation of the moiré pattern in bilayer $WS_2$ is more energetically favorable for low-angle GB. A zoom-in view of the atomic structure of the GB reveals how the bilayer stacking orientation develops from AB to a moiré pattern without a lattice discontinuity (Figures 6e, S13 and S14). Again, an *in-plan*e translation of the bilayer is observed in the high-resolution STEM image. Here, several competitive factors, i.e., weak interlayer van der Waals forces and intralayer GB results in a buildup strain in the bilayer structure.

## CONCLUSIONS

In summary, we have developed a PLD-based synthesis of quasi-continuous mono- and multilayers $WS_2$ films based on sulfurization of $WO_{3-x}$ epitaxial films. We have shown that the composition of $WO_{3-x}$ films determines the morphology and optical properties of $WS_2$ monolayers synthesizes in a CVD process. The presence of oxygen vacancies in tungsten oxide film is highly favorable for successful film conversion. The optimized oxide growth conditions (800 $^{o}$C deposition temperature using Ar background pressure) results in formation of a stoichiometric 2H-$WS_2$ monolayer film with a large size of crystals, reaching 26 um on the films edges.

## METHODS

### Synthesis of $WS_2$ crystals

A customized pulsed laser deposition (PLD) setup consisting of a target carousel equipped with a water cooled baffle (from McAllister Technical Services) and an IR lamp substrate heating system (from *AJA* International INC) was used to first produce a variety of tungsten oxide precursors which were then sulfurized for conversion to $WS_2$ crystals. The cooling of the target is not critical in this work, but from our experience it is strictly necessary for ablation of sulfide materials.

*Growth of precursor epitaxial WOx films by pulsed laser deposition (PLD).* The thin-films were grown by laser ablation of $WO_3$ target (1 inch diameter, 6 mm thickness, 99.95% purity, from Testbourne Ltd.) using a 248 nm KrF laser (pulse duration $\tau$=20 ns) operating at laser frequency of 1 Hz. The target was placed at a distance of ~7 cm from the substrate. Prior growth, the target was ablated using a number of 30 laser pulses. During deposition, the laser beam was rastered across the target using motorized mirrors and the target was rotated at the angular speed of 6 rpm to ensure uniform ablation and consequently films with a good thickness uniformity. The laser fluence on the target was maintained at ~ 1.5 J/cm$^2$. Prior deposition, the vacuum chamber was pumped down to base pressure of ~$5 \times 10^{-7}$ mbar and background gas was introduced manually using a needle valve. To growth of oxide precursors was carried out O (99.9995% purity, Air Liquide Danmark A/S) or Ar (99.9999% purity, Air Liquide Danmark A/S). In all experiments, the gas pressure was maintained at $5 \times 10^{-3}$ mbar. The films were grown (0001) $Al_2O_3$ and $SiO_2$ (300 nm)/Si substrates of 1 x 1 cm$^2$ in size cut from large wafers. Before loading the substrates into the PLD chamber, they were sequentially cleaned in acetone, isopropanol (IPA) and deionized water (DI) water for 10 min in each solvent combined with sonication. The substrate were heated using a ramp rate of 15 ºC/min to the desired growth temperature and finally cooled down to room temperature using a similar cooling rate of 15 ºC/min. The $WO_x$ precursors were grown on a temperature range varying from 500 ºC to 800 ºC using a number of ~40 laser pulses. A minimum number of 30 pulses was required to obtain any deposition.

*Sulfurization of WOx films***.** The sulfurization of $WO_x$ precursor films was performed as follows. (1) Firstly, the precursor films were placed facing up onto a flat graphite plate and loaded in the middle of the three-zone tube furnace of a 3 inch outside diameter. A narrow alumina ceramic boat containing ~2.0 g of sulfur flakes (≥99.99 % purity, from Sigma Aldrich) was placed outside the main heating area of the quartz tube and heated independently using an external heater. The distance between the ceramic boat and the samples was 35 cm. (2) The system was pumped down to $5 \times 10^{-2}$ mbar and sequentially flushed 5 times by pressuring to 500 mbar with Ar-$H_2$ mixture (5 % volume fraction, 99.9995 % purity, Air Liquide Danmark A/S) to remove residual gases, such as oxygen and water vapor. Then, a 200 sccm Ar-5 % $H_2$ gas flow was introduced as a carrier gas throughout the entire process. The tube furnace was then heated at a rate of 15 ºC/min to 900 ºC and the sulfurization process took place at 900 ºC for 30 min. Moreover, when the furnace temperature reached 900 ºC, the external heater was set to 250 ºC to produce an adequate sulfur evaporation rate. The system was finally cooled down at a rate of 15 ºC/min to 500 ºC, and then naturally to room temperature.

**Structural characterization of the $WS_2$ films**

*Raman/PL spectroscopy.* Raman and PL spectra were collected using home-built Raman confocal spectroscopy setup using a 532 nm excitation laser. The samples were illuminated through a 40x objective lens (NA=0.75), type Nikon CFI Plan Flour. The spectrometer is a Spectra Pro HRS-750 scanning monochromator from Princeton Instruments equipped with suitable gratings of 300 gr/mm (for PL measurements) and 1800 gr/mm (for Raman acquisition) and a cryogenically-cooled, ultra-low noise Pylon CCD camera, type PyLoN:100BR from Princeton Instruments. The wavelength calibration was carried using an Ar-Ne light source mounted directly on a side entrance slit of the spectrometer. The spectrometer resolution was found to be ~1 $cm^{-1}$ at 532 nm. Raman spectra were collected using various integration time and then normalized. The PL spectra were taken at a low laser power below 10 mW to prevent overheating.
*Scanning Electron Microscopy (SEM).* The morphologies of the $WS_2$ crystals grown on either (0001) sapphire or $SiO_2$/Si substrates were analyzed using a Zeiss Merlin microscope operated using an In Lens detector, acceleration voltages in the 1−2 kV and working distance between 3 and 4 cm.
*Atomic Force Microscopy (AFM).* AFM measurements were conducted at room temperature using a Bruker Icon AFM microscope.
*X-ray Photoelectron Spectroscopy (XPS).* The XPS spectra were recorded using a Thermo Fisher Nexsa XPS system with Al Kα 1486.6 eV excitation source. The energy scale of the XPS was calibrated using Au $4f_{7/2}$ line centered at 84.0 eV, and the line positions were calibrated using $W4f_{7/2}$ peak of WO3 at 36.0 eV, which is present in all the samples.

*Scanning Transmission Electron Microscopy (STEM).* The $WS_2$ specimens were transferred onto a TEM grid with the aid of cellulose acetate butyrate (CAB) coating followed by KOH etching. CAB is used because of its hydrophobic nature compared to the hydrophilic substrate. In this transfer process the CAB solution (20 mg/100 ml in ethyl acetate) produces 15 µm-thick polymer. The specimens on sapphire were spin coated with CAB using a 1500 rpm rotation speed followed by heating for few minutes on a hot plate. The samples were dipped inside KOH solution in order to etch them from sapphire and washed several times with DI water for the complete removal of KOH. The samples were then transferred to TEM grid and kept them in air for several minutes for drying. The presence of the samples before and after cleaning with acetone were confirmed by Raman measurement. Afterwards, the samples were cleaned with acetone to remove CAB on the samples. The samples transferred on the TEM grid were baked at a temperature of 160 °C in vacuum for 8 h before loading them in the STEM chamber for imaging. The STEM measurements were performed using an aberration-corrected Nion UltraSTEM 100 microscope equipped with a cold field emission gun. The images were acquired at an acceleration voltage of 60 kV and a semiconvergence angle of 31 mrad.

*DFT Calculations.* The DFT calculations were performed within the projected-augmented (PAW) method implemented in the electronic structure code GPAW (21). We used a plane wave basis set with a cutoff of 800 eV and included exchange and correlations with the Perdew-Burke-Enzerhof (PBE) functional (20). Reciprocal space was sampled with a *k*-point density of 6 $Å^{-1}$ including the Γ point. The atomic positions were relaxed at a fixed cell until forces were less than $10^{-2}$ $eVÅ^{-1}$.

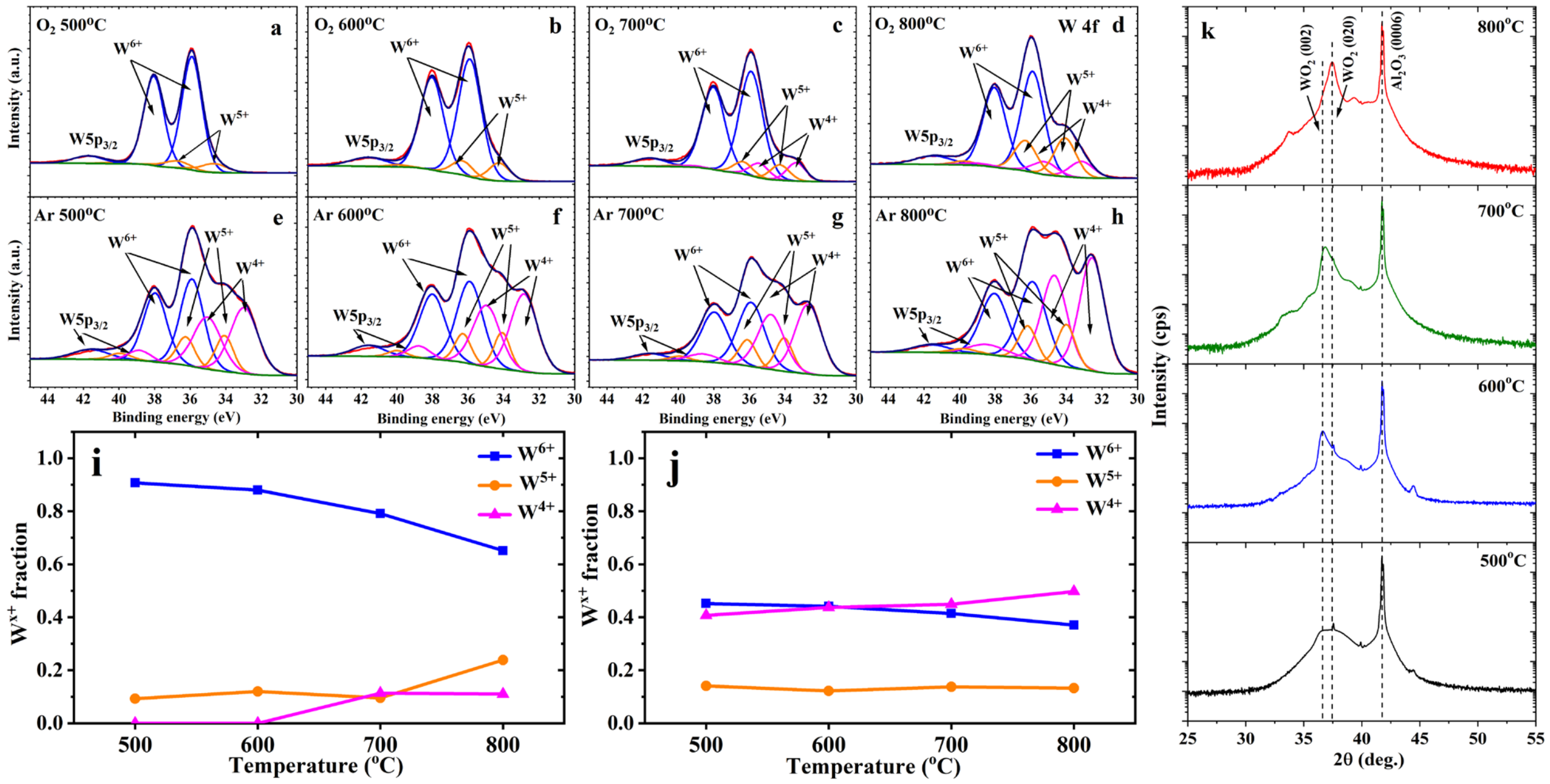

**Figure 1. Chemical composition of PLD-grown $WO_{3-x}$ precursors.** W4f core level XPS spectra of $WO_{3-x}$ films grown in temperature range from 500 ºC to 800 ºC in $O_2$ (a-d) and Ar (e-h). (i, j) The corresponding $W^{x+}$ fraction extracted from the XPS spectra. The XPS spectra were fitted with peaks attributed to three sets of doublets corresponding to $W^{6+}$ (blue lines), $W^{5+}$(orange lines) and $W^{4+}$ (magenta lines) valence states, and their corresponding $W5p_{3/2}$ (same color as a doublet). All the films were grown on (0001) $Al_2O_3$ substrates using 40 laser pulses. (k) XRD θ-2θ patterns of $WO_2$ films grown on (0001) $Al_2O_3$ substrate in Ar in a temperature range from 500 $^0$C to 800 $^0$C.

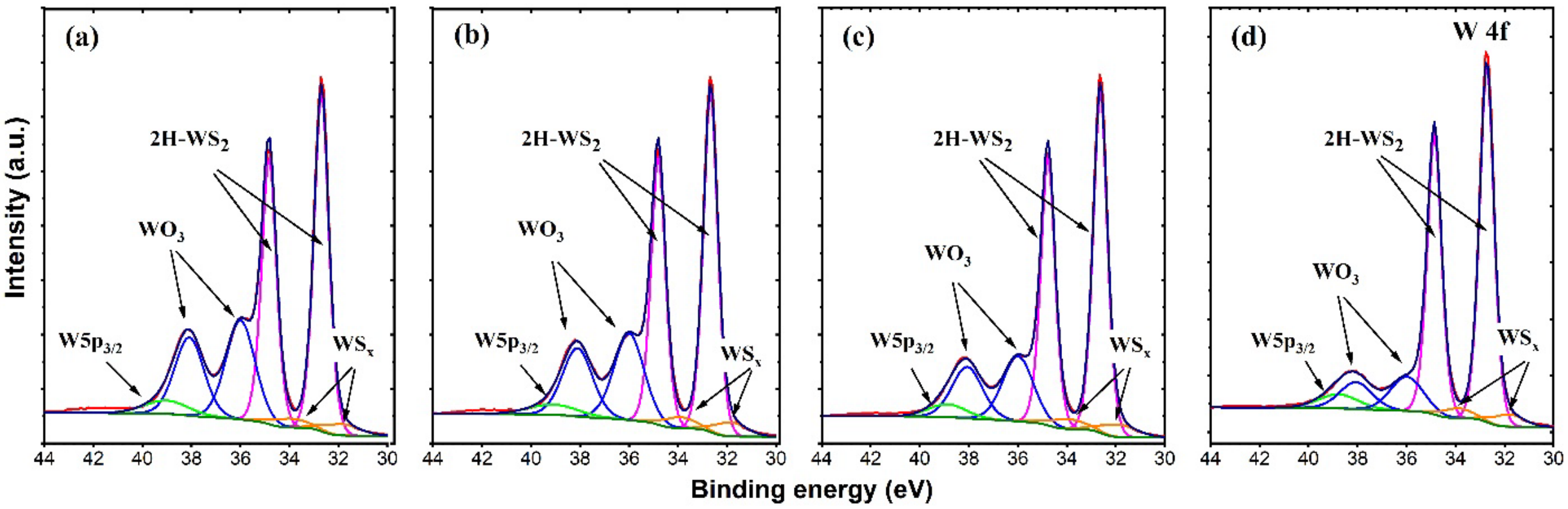

**Figure 2. Chemical analysis of $WS_2$ using XPS.** W 4f core level XPS spectra of $WS_2$ films obtained upon sulfurization of $WO_{3-x}$ precursors grown in Ar at 500 ºC (a), 600 ºC (b), 700 ºC (c) and 800 ºC (d). The XPS spectra peaks were attributed to three sets of doublets corresponding to $W^{6+}$ (blue lines), $W^{4+}$ (magenta lines) and $W^{x+}$(orange lines) valence states, and their corresponding W $5p_{3/2}$ (light green). $WS_x$ denotes S-deficient $WS_2$.

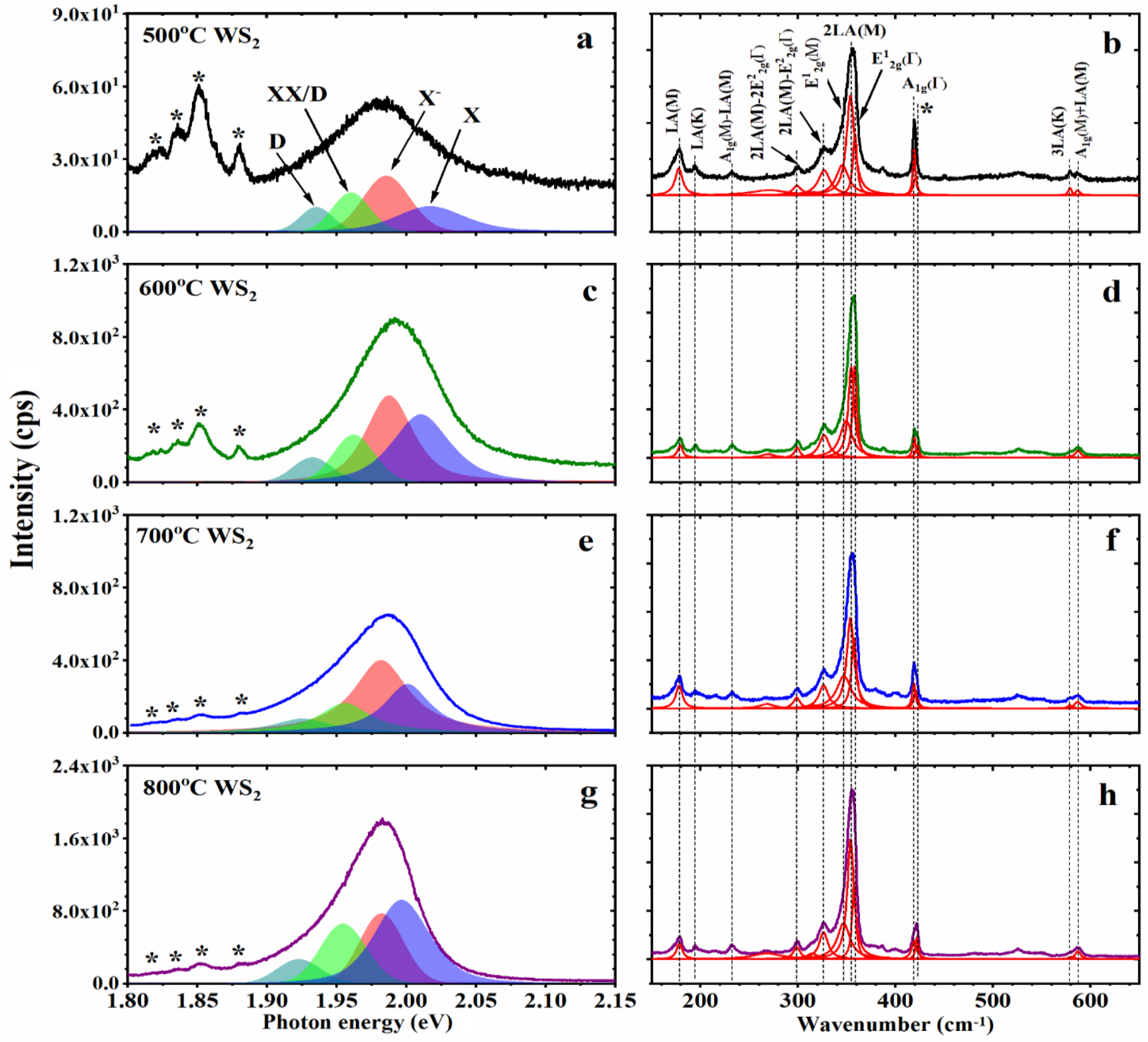

**Figure 3. Optical properties of WS2**. PL and Raman spectra of $WS_2$ obtained by sulfurization at 900 ºC of WO3-xprecursors grown in Ar at a temperature of 500 ºC (a, b), 600 ºC (c, d), 700 ºC (e, f), 800 ºC (g, h). Deconvolution of the PL spectra using Lorentzian multiple-peak fitting into four peaks corresponding to neutral exciton (X) (blue), negative exciton – trion ($X^{-}$) (red), bi-exciton (XX) (green), and defected-mediated exciton (D) (dark cyan). Deconvolution of the Raman spectra using Lorentzian multiple-peak fitting is marked with red curves (major contributions presented).

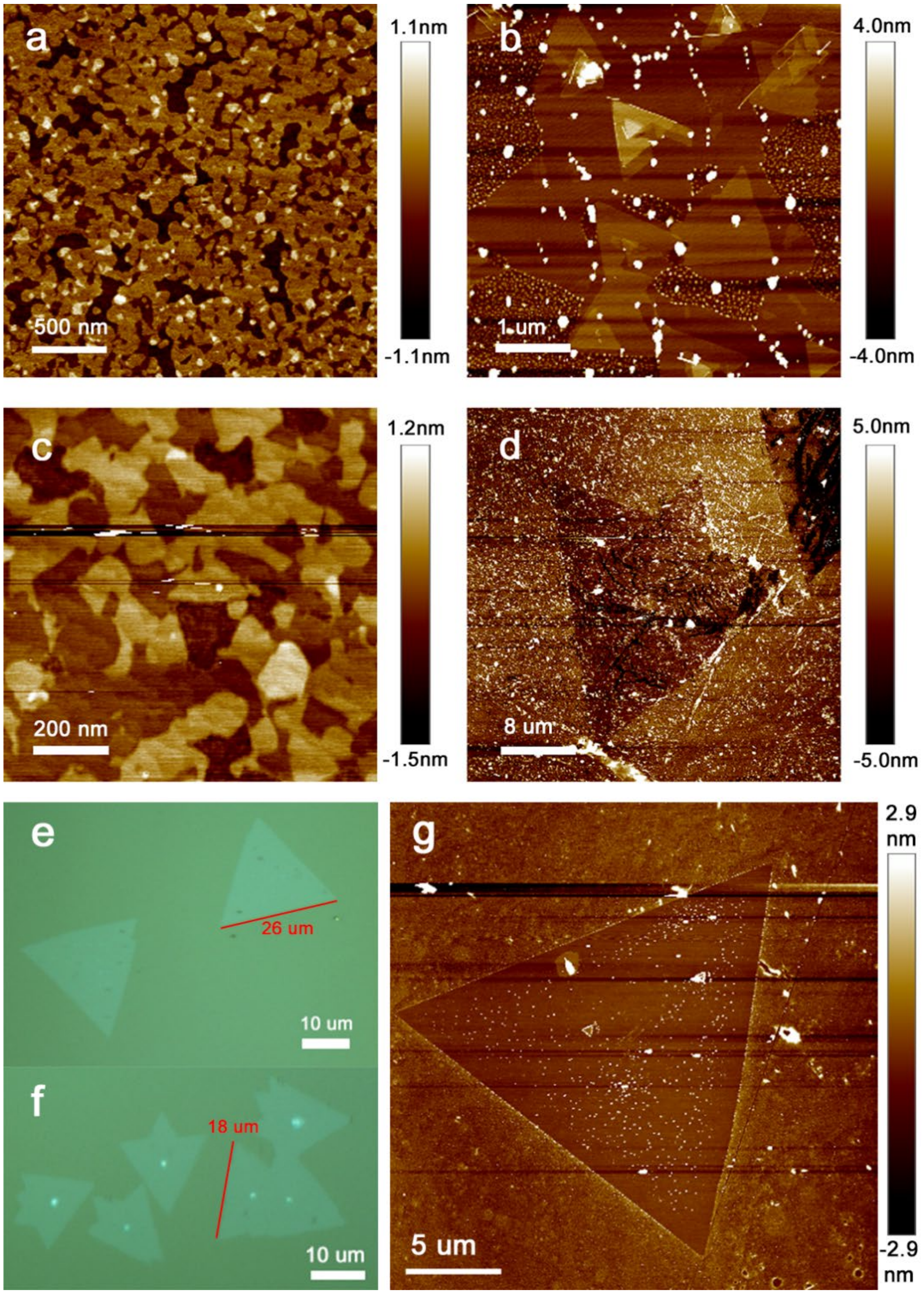

**Figure 4**. **Morphology of $WS_2$ crystals**. AFM images showing the change in the morphology of the of $WS_2$ as a function of the growth parameters of the $WO_{3-x}$ precursors: (a) 500 ºC, $O_2$; (b) 800 ºC, $O_2$; (c) 500 ºC, Ar; (d) 800 ºC, Ar, (e, f) Optical images of $WS_2$ monolayers based on 800 ºC-grown precursors, and (g) the corresponding AFM image.

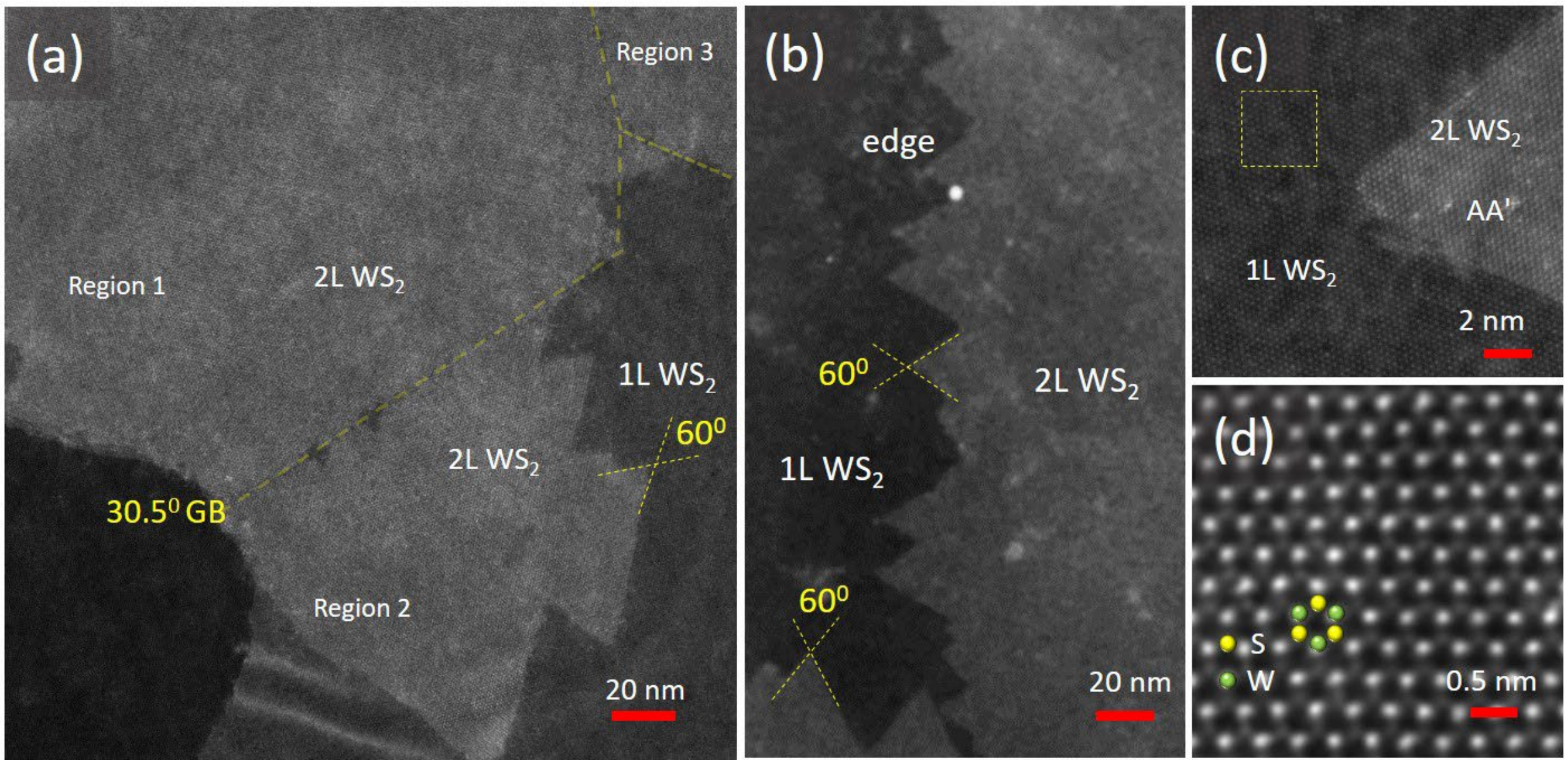

**Figure 5**. **Atomic structure of mono-bilayer $WS_2$.** ADF-STEM images of mono-bilayer $WS_2$ derived from $500^0$C (a) and $800^0$C-grown precursors (b). (c) Magnified STEM image showing of triangular-shape edges of bilayer $WS_2$ with a 2H (AA$^{'}$) interface. (d) Magnified view of the yellow box in (c) showing the atomic resolution image of pristine monolayer.

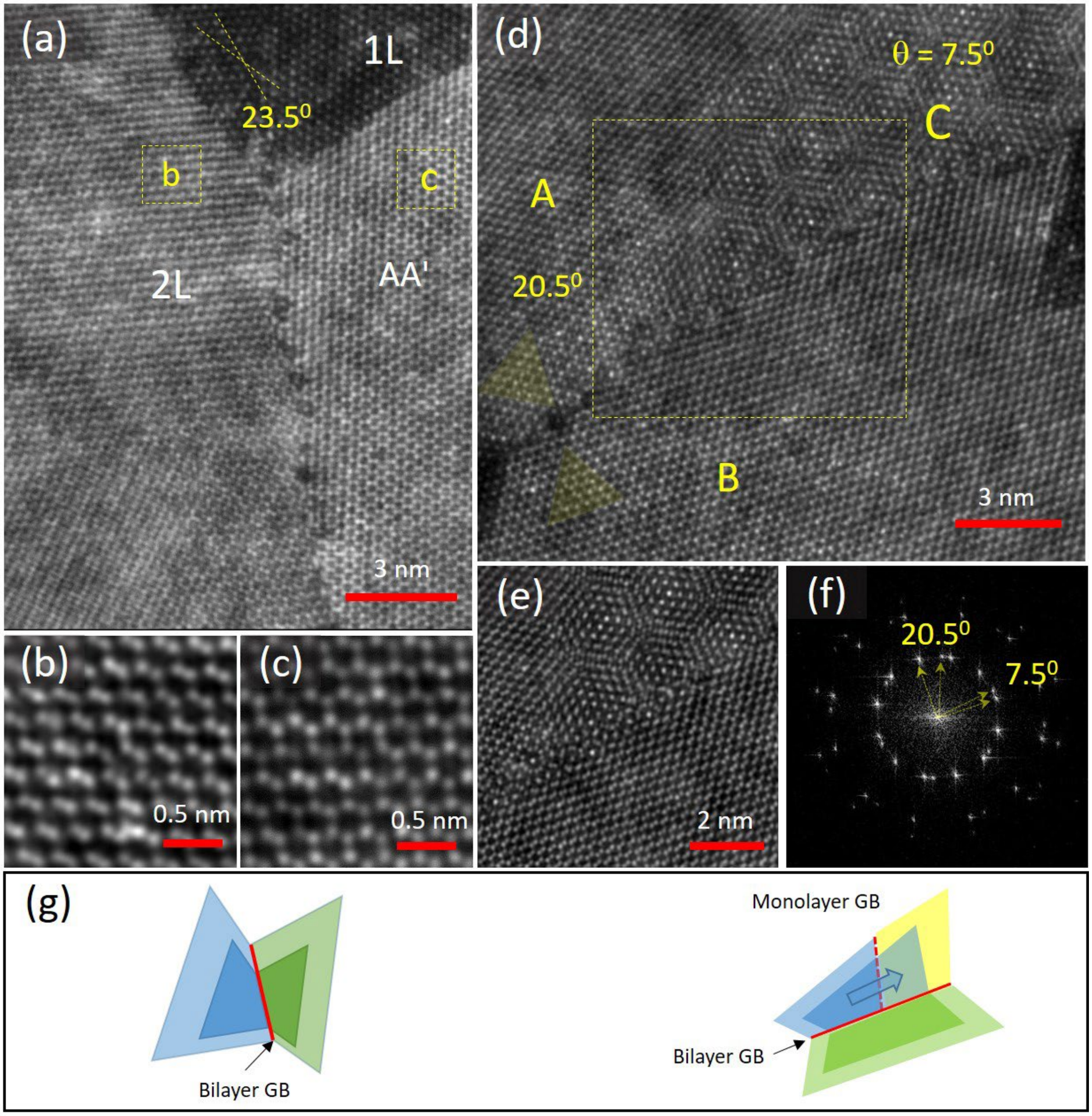

**Figure 6**. **Atomic structure of bilayer $WS_2$. a)** ADF-STEM images of $23.5^0$ bilayer GB showing bilayer $WS_2$ sharing the same GB as the monolayer counterpart. Panels b) and c) are zoom-in views of bilayer domains from regions highlighted in (a). (d) STEM image of $20.5^0$ bilayer GB showing AA$^{\prime}$/AB orientation and moiré pattern. e) High-resolution image of bilayer domains taken roughly on the region highlighted in d). (f) Fast Fourier transform (FFT) of the AF STEM image shown in (d). (g) Left- and right-hand illustrations revealing the formation of bilayer $WS_2$ in (a), and (d), respectively.

**Table 1**. The average composition of the $WO_{3-x}$ precursors and $WS_2$ derived from XPS.

| **Background gas ($8\times10^{-3}$ mbar)** | **Substrate temperature (ºC)** | **Oxide Precursors** | **Sulfides** | | | |
|---|---|---|---|---|---|---|
| | | $WO_{3-x}$ Stoichiometry(*) | $WS_2$ Stoichiometry | $WS_x$ Stoichiometry | Residual $WO_3$ | Residual $WS_x$ |
| **Oxygen** | 500 ºC | 2.95 | - | - | - | - |
| | 600 ºC | 2.94 | - | - | - | - |
| | 700 ºC | 2.84 | 2.10 | 2.10 | 0.50 | 0.09 |
| | 800 ºC | 2.76 | 2.15 | 2.15 | 0.42 | 0.05 |
| **Argon** | 500 ºC | 2.52 | 2.00 | 2.00 | 0.34 | 0.09 |
| | 600 ºC | 2.50 | 2.00 | 2.00 | 0.33 | 0.09 |
| | 700 ºC | 2.48 | 2.00 | 2.00 | 0.27 | 0.08 |
| | 800 ºC | 2.44 | 2.02 | 2.02 | 0.20 | 0.06 |

(*) Average composition of the films accounting all various oxidation states of W in $WO_{3-x}$. The $WS_2$ stoichiometry was estimated by taking into account contributions from both 2H-$WS_2$ and $WS_x$.

**Table 2.** Raman peak positions, intensity ratios and FWHM for the $WS_2$ layers grown using different precursors. The excitation wavelength was 532 nm. The data were fitted using Lorentzian functions.

| **Background gas ($8\times10^{-3}$ mbar)** | **Substrate temperature ($^0$C)** | **Sulfides** | | | | |
|---|---|---|---|---|---|---|
| | | $A_{1g}(\Gamma)$ ($cm^{-1}$) | FWHM $A_{1g}(\Gamma)$ ($cm^{-1}$) | $E^1_{2g}(\Gamma)$ ($cm^{-1}$) | $\Delta k$ ($cm^{-1}$) | 2LA(M) ($cm^{-1}$) |
| **Oxygen** | 500 ºC | - | - | - | - | |
| | 600 ºC | - | - | - | - | |
| | 700 ºC | 419.2 | 3.84 | 358.4 | 61.2 | 178.8 |
| | 800 ºC | 419.0 | 3.38 | 358.5 | 60.5 | 178.8 |
| **Argon** | 500 ºC | 419.2 | 3.08 | 358.5 | 60.7 | 177.9 |
| | 600 ºC | 419.4 | 3.66 | 358.3 | 61.1 | 179.1 |
| | 700 ºC | 418.8 | 3.90 | 357.5 | 61.3 | 178.2 |
| | 800 ºC | 419.1 | 6.39 | 357.5 | 61.6 | 178.5 |

ASSOCIATED CONTENT

**Supporting Information**. A listing of the contents of each file supplied as Supporting Information should be included. For instructions on what should be included in the Supporting Information as well as how to prepare this material for publications, refer to the journal's Instructions for Authors. The following files are available free of charge. Table S1, XPS data of $WO_{3-x}$; Table S2, XPS data of $WS_2$; Figure S1, S 2p core level XPS spectra of $WS_2$; Figure S2, S 2p core level XPS spectra of $WS_2$; Figure S3, W 4f core level XPS spectra of $WS_2$ ; Figure S4, $\theta - 2\theta$ XRD pattern of $WO_2$ on (0001) $Al_2O_3$; Figure S5, High-resolution SEM images of $WS_2$; Figure S6, Optical images of $WS_2$ monolayers; Section Raman and PL studies; Figure S7, DFT-calculated formation energy ($E^F$) of oxygen vacancies in $WO_{3-x}$ oxides; Figure S8, Equilibrium oxygen vacancy concentration as a function of temperature; Figure S9, Visualization of sulfur interstitial ($S_i$) and sulfur on oxygen site ($S_O$) in $WO_3$; Figure S10, Visualization of systems in the slab calculations; Figure S11, Intrinsic point defects on monolayer $WS_2$; Figure S12, STEM image of $20.5^0$ grain boundary in monolayer $WS_2$; Figure S13, Low-resolution STEM image showing a $30.5^0$ monolayer; Figure S14, STEM and Fourier-filtered images of bilayer $WS_2$.

AUTHOR INFORMATION

**Corresponding Author**

Stela Canulescu − Department of Photonics Engineering, Technical University of Denmark,

4000 Roskilde, Denmark; orcid.org/0000-0003-3786-2598; Email: stec@fotonik.dtu.dk.

**Present Addresses**

NA

**Author Contributions**

The manuscript was written through contributions of all authors. All authors have given approval to the final version of the manuscript.

**Funding Sources**

NA

**Notes**

Any additional relevant notes should be placed here.

ACKNOWLEDGMENT

S.C. acknowledges support from the Independent Research Fund Denmark, Sapere Aude grant (project number 8049- 00095B). K.S.T. acknowledges support from the Center for Nanostructured Graphene (CNG) under the Danish National Research Foundation (project DNRF103) and from the European Research Council (ERC) under the European Union's Horizon 2020 research and innovation program (ERC grant no. 773122, LIMA). STEM imaging was conducted as a part of a user project at the Center for Nanophase Materials Sciences, which is a DOE Office of Science User Facility.

ABBREVIATIONS

PLD pulsed laser deposition